\documentclass[12pt,preprint]{aastex}

\begin{document}

\title{Triggering Collapse of the Presolar Dense Cloud Core 
and Injecting Short-Lived Radioisotopes with a Shock Wave. 
V. Nonisothermal Collapse Regime}

\author{Alan P.~Boss}
\affil{Department of Terrestrial Magnetism, Carnegie Institution for
Science, 5241 Broad Branch Road, NW, Washington, DC 20015-1305}
\email{aboss@carnegiescience.edu}

\begin{abstract}

 Recent meteoritical analyses support an initial abundance of the
short-lived radioisotope $^{60}$Fe that may be high enough to require
nucleosynthesis in a core collapse supernova, followed by rapid
incorporation into primitive meteoritical components, rather than a
scenario where such isotopes were inherited from a well-mixed region
of a giant molecular cloud polluted by a variety of supernovae remnants
and massive star winds. This paper continues to explore the former scenario, 
by calculating three dimensional, adaptive mesh refinement, hydrodynamical
code (FLASH 2.5) models of the self-gravitational, dynamical
collapse of a molecular cloud core that has been struck by a thin shock
front with a speed of 40 km/sec, leading to the injection of shock front
matter into the collapsing cloud through the formation of Rayleigh-Taylor
fingers at the shock-cloud intersection. These models extend the previous
work into the nonisothermal collapse regime using a polytropic approximation 
to represent compressional heating in the optically thick protostar. The
models show that the injection efficiencies of shock front material
are enhanced compared to previous models, which were not carried into
the nonisothermal regime and so did not reach such high densities. The
new models, combined with the recent estimates of initial $^{60}$Fe
abundances, imply that the supernova triggering and injection scenario
remains as a plausible explanation for the origin of the short-lived 
radioisotopes involved in the formation of our solar system.

\end{abstract}


\section{Introduction}

 The discovery of reproducible evidence for live $^{26}$Al (Lee et al. 1976), 
with a half-life of 0.72 Myr, in Ca, Al-rich refractory inclusions (CAIs) 
from the Allende meteorite led directly to the hypothesis of the 
injection of core-collapse supernova-derived (CCSN)
short-lived radioisotopes (SLRIs) such $^{26}$Al (as well as $^{60}$Fe) into 
the presolar cloud and the triggering of the collapse of the cloud to form
the solar nebula (Cameron \& Truran 1977). However, $^{26}$Al is also produced
in abundance during the Wolf-Rayet (WR) star phase, prior to the supernova explosion
of massive stars (e.g., Young 2016), whereas $^{60}$Fe, with a half-life of 2.6 Myr,
is only produced in significant amounts by CCSN and AGB stars. As a result,
a high initial level of $^{60}$Fe is often considered the {\it smoking gun} for the 
supernova triggering and injection hypothesis. 

 Early estimates of the initial amount of $^{60}$Fe in chondrules from 
ordinary chondritic (OC) meteorites implied a ratio of $^{60}$Fe/$^{56}$Fe
$\sim 5-10 \times 10^{-7}$ (Tachibana et al. 2006), high enough to require
a CCSN for its nucleosynthesis shortly before its incorporation into the 
earliest solids in the solar nebula. However, subsequent analyses of
bulk samples of different meteorite types produced a much lower initial 
ratio of $^{60}$Fe/$^{56}$Fe $\sim 1.15 \times 10^{-8}$ (Tang \& Dauphas 2012),
seemingly refuting the smoking gun theory. Using that lower ratio, Young 
(2014, 2016) argued that there was no need for a recent SN injection, and
that the initial abundances of a dozen SLRIs could be explained by the
contributions of massive star winds to the general interstellar medium
in regions of star formation. However, other studies continued to find
high initial $^{60}$Fe/$^{56}$Fe ratios, now in unequilibrated OC (UOC), 
of $\sim 7 \times 10^{-7}$, $\sim 2 - 8 \times 10^{-7}$, and
$\sim 8 - 11 \times 10^{-7}$, respectively (Mishra \& Goswami 2014; 
Mishra \& Chaussidon 2014; Mishra et al. 2016). As a result, a mystery 
arose regarding $^{60}$Fe analyses of individual chondrules compared to 
bulk samples.

 The puzzling situation with regard to $^{60}$Fe seems to have been resolved by 
the work of Telus et al. (2016). Telus et al. (2016) found that the bulk sample
estimates were skewed toward low initial $^{60}$Fe/$^{56}$Fe ratios because of 
fluid transport of Fe and Ni along chondrule fractures during aqueous alteration 
on the parent body or during terrestrial weathering. While the initial
$^{60}$Fe/$^{56}$Fe value is still somewhat uncertain, Telus et al. (2017)
have found that UOCs appear to have levels of $^{60}$Ni consistent with
initial ratios of $^{60}$Fe/$^{56}$Fe as high as $5.6 \times 10^{-7}$. This 
value supports a CCSN supernova as a plausible source for many of the SLRIs, and 
refutes the scenario advanced by Young (2014, 2016), at least regarding $^{60}$Fe.
Furthermore, CCSN nucleosynthesis models by Banerjee et al. (2016) have shown 
that neutrino spallation reactions can produce the requisite amount of live 
$^{10}$Be (half-life of 1.4 Myr) found in certain CAIs
in a supernova that occurred about 1 Myr prior to CAI formation. Given
these recent cosmochemical results, the supernova trigger hypothesis continues
to be worthy of detailed investigation.

 One variation on the shock-triggered collapse scenario is the injection
of CCSN-derived material directly into an already-formed solar nebula
(e.g., Oullette et al. 2007, 2010). Parker \& Dale (2016) concluded that 
such a scenario is probably more likely to occur than the original scenario
of injection into a pre-collapse cloud core, while Nicolson \& Parker (2017)
found that disk enrichment was as likely to occur in low mass star clusters
as in more massive clusters. Given that we only know about the presence 
of possibly high initial levels of certain SLRIs in meteorites from
our own solar system [cf., Jura et al. (2013) for a different point 
of view, based on observations of metal-rich white dwarf stars], 
it is unclear how one can use probabilistic arguments 
to constrain the formation of a single planetary system in a galaxy that
contains billions of planetary systems (cf. Lichtenberg et al. 2016).
The crucial question is whether the scenario is {\it possible}, rather 
than if it is {\it probable}, given our sample of one.  
Oullette et al. (2007, 2010) studied SN shock wave injection
into the solar nebula, and found that only solid particles larger
than $\sim$ 1 $\mu$m can be injected; smaller particles and gas in the SN shock
front are unable to penetrate into the much higher density protoplanetary disk.
Goodson et al. (2016) performed a similar study of particle injection, but 
into a molecular cloud rather than a disk, and also found that particles larger 
than $\sim$ 1 $\mu$m were preferentially injected deep with the target cloud.
However, analysis of Al and Fe dust grains in SN ejecta constrain their 
sizes to be less than 0.01 $\mu$m (Bocchio et al. 2016), effectively 
severely limiting the value of both of these injection mechanism scenarios. 

 Vasileiadis et al. (2013) and Kuffmeier et al. (2016) have modeled the
evolution of a $\sim 10^5 M_\odot$ Giant Molecular Cloud (GMC), and 
shown how the SLRI’s ejected by CCSN in the GMC can pollute the surrounding 
gas to highly variable levels, ranging from $^{26}$Al/$^{27}$Al 
$\sim 10^{-7}$ to $\sim 10^{-1}$, much higher than the canonical solar system 
ratio of $^{26}$Al/$^{27}$Al $\sim 5 \times 10^{-5}$ (Lee et al. 1976).
Ratios of $^{60}$Fe/$^{56}$Fe were as high $\sim 10^{-3}$ (Kuffmeier et al. 
2016), again much higher than any meteoritical estimates.
Given that the GMC simulation covered a 3D region 40 pc in extent, with
computational cell densities no higher than $\sim 10^{-16}$g cm$^{-3}$,
data sampled every 0.05 Myr, and minimum grid spacings of 126 AU, this 
ambitious simulation was unable to follow in detail the interactions of
specific supernova shock fronts with any dense molecular cloud cores.
Such interactions are the focus of the present series of studies,
where SN shocks with $\sim$ 60 AU thicknesses strike clouds and within 
0.1 Myr trigger collapse to densities of over $\sim 10^{-11}$ g cm$^{-3}$.
In fact, star formation in the Vasileiadis et al. (2013) and Kuffmeier 
et al. (2016) GMC calculation is not the result of SN shock 
wave triggering, but rather the eventual self-gravitational collapse of
pre-stellar cores formed by GMC turbulence over 4 Myr of GMC evolution.

 Balazs et al. (2004) presented observational evidence that a supernova 
shock front triggered the formation of T Tauri stars in the L1251 cloud.
As reviewed by Boss (2016), the numerical study of the shock-triggered 
collapse and injection scenario for the origin of the solar system SLRIs 
began with the relatively crude three dimensional (3D) models of Boss (1995).
Subsequent theoretical models retreated to the higher spatial resolution
afforded by axisymmetric, two dimensional (2D) models (e.g., Foster \& Boss 
1996; Vanhala \& Boss 2000), which revealed the role of Rayleigh-Taylor (RT) 
fingers in the injection mechanism. Boss et al. (2008, 2010)
introduced the use of the FLASH adaptive mesh refinement (AMR) code in
order to better resolve the thin shock fronts and RT fingers with a reasonable
number of spatial grid points, and FLASH has been
the basis of our modeling ever since. Boss \& Keiser (2013) used 2D FLASH AMR
models to study rotating presolar clouds that collapsed to form disks.
Boss \& Keiser (2014, 2015) extended those models to 3D, with rotational
axes either parallel to, or perpendicular to, the direction of propagation of the
SN shock front, respectively. Li et al. (2014) and  Falle et al. (2017)
used AMR codes to study the triggering of collapse by isothermal shocks 
striking non-rotating, isothermal, Bonnor-Ebert spheres.

 All of the previous FLASH AMR code models have been restricted
to optically thin regimes ($< 10^{-13}$ g cm$^{-3}$) where
molecular line cooling by H$_2$O, CO, and H$_2$ are effective at cooling the
shock-compressed regions (Boss et al. 2008, 2010; Boss \& Keiser 2013, 2014, 2015).
The Vasileiadis et al. (2013) and Kuffmeier et al. (2016, 2017) GMC 
calculations have a similar limitation to optically thin regimes. In our case, 
this restriction allows one to learn if a given target cloud can be shocked into
collapse, but it prohibits following the calculation to much later times, after
the central protostar forms, surrounded by infalling, rotating gas and dust
that will form the solar nebula. To a first approximation, nonisothermal effects 
in optically thick regimes can be approximated by the use of a simple 
polytropic ($\gamma = 7/5$) pressure law, mimicking the compressional 
heating in molecular hydrogen encountered at 
densities above $\sim 10^{-13}$ g cm$^{-3}$ (e.g., Boss 1980). Adding
this new handling of the equation of state in the nonisothermal regime
is the motivation for the 3D models presented in this paper. 

\section{Numerical Hydrodynamics Code}

 As with the previous papers in this series, the models were calculated 
with the FLASH2.5 AMR hydrodynamics code (Fryxell et al. 2000). Full details 
about the implementation of FLASH 2.5 may be found in Boss et al. (2010). 
We restrict ourselves here to noting what has been changed in the present
models compared to the most recent previous 3D models (i.e., Boss \& Keiser 
2014, 2015). As before, a color field is used to follow the transport of
the gas and dust initially residing in the shock front.

 Initially, the low density target cloud and the surrounding gas are isothermal 
at 10 K, whereas the SN shock front and post-shock gas are isothermal at 1000 K. 
The FLASH code follows the subsequent compressional heating as well as 
cooling of the interaction between the shock and the cloud by the molecular 
species H$_2$O, CO, and H$_2$, resulting in a radiative cooling rate appropriate 
for rotational and vibrational transitions of optically thin, warm 
molecular gas (Boss et al. 2010). As a result, the shock front temperatures
are unchanged from those in the previous simulations.
The primary change is then the use of a 
polytropic equation of state when densities high enough to produce an optically 
thick cloud core arise, i.e., for densities above some critical density 
$\rho_{cr} \sim 10^{-13}$ g cm$^{-3}$ (Boss 1980). Molecular gas cooling is then
prohibited for densities greater than $\rho_{cr}$, when the gas pressure 
$p$ depends on the gas density $\rho$ as $p \propto \rho^\gamma$, where the
polytropic exponent $\gamma$ equals 7/5 for molecular hydrogen gas.
We tested several different
variations in $\rho_{cr}$ in order to find the value that was best able to
reproduce the dependence of the central temperature on the central density
found in the spherically symmetrical collapse models of Vaytet et al. (2013),
who performed detailed multigroup radiative hydrodynamics calculations of
the collapse of solar-mass uniform density spheres.

\section{Initial Conditions}

 Table 1 lists the variations in the initial conditions for the models,
which are all variations on two rotating models considered previously in this
series (e.g., Boss \& Keiser 2014): a 2.2 $M_\odot$ cloud with a radius 
of 0.053 pc and a Bonnor-Ebert radial density profile, with
solid body rotation about the $\hat y$ axis (the direction of propagation
of the shock wave) at an angular frequency of $\Omega_i = 10^{-14}$ or 
$10^{-13}$ rad s$^{-1}$. The shock wave propagates toward 
the $- \hat y$ direction with a shock speed of 40 km s$^{-1}$ with an initial 
shock width of $3 \times 10^{-4}$ pc and an initial shock density of 
$7.2 \times 10^{-18}$ g cm$^{-3}$. In order to test the sensitivity of
the results to the presence of noisy initial conditions, two models 
were run with random noise in the interval of [0,1] inserted in the
ambient gas surrounding the uniform density target cloud, resulting in 
the ambient medium having density variations ranging from a maximum density 
of $1.2 \times 10^{-20}$ g cm$^{-3}$ to a minimum density 40 times lower,
$3 \times 10^{-22}$ g cm$^{-3}$.

 Boss et al. (2010) studied the effect of shock wave speeds ranging 
from 1 km s$^{-1}$ to 100 km s$^{-1}$, finding that sustained triggered 
collapse was only possible for shock speeds in the range of 5 km s$^{-1}$ 
to 70 km s$^{-1}$: slower shocks failed to inject shock wave material, 
while faster shocks shredded the clouds and did not trigger collapse.
Boss \& Keiser (2010) presented results for shock speeds of 40 km s$^{-1}$
and varied shock thicknesses, as high as 100 times thicker than used here, 
in order to distinguish between relatively thin SN shocks and
relatively thick planetary nebula winds from AGB stars. They found 
that 40 km/s shocks thicker than those considered here shredded the 
target clouds without inducing sustained collapse, even when the 
shock densities were lower than assumed here, and so concluded 
that an AGB star was an unlikely source of the solar system SLRIs.
Boss et al. (2010) noted out that these simulations are meant to 
represent the radiative phase of SN shock wave evolution, which occurs 
after the Sedov blast wave phase (Chevalier 1974). In the radiative 
phase, the shock sweeps up gas and dust while radiatively cooling, 
as is modeled here.

 The FLASH 2.5 code AMR numerical grids start with 6 grid blocks
along the $\hat x$ and $\hat z$ directions, and either 9 or 12 grid blocks
in the $\hat y$ direction (the latter for the six models with 
$y_{min} = -2.0 \times 10^{17}$ cm), with each grid block consisting of $8^3$ cells.
Models started with either 3 or 4 levels of refinement (with each level refined by 
a factor of two) on the initial grid blocks. The number of levels of grid refinement 
was increased during the evolutions to as many as 8 levels when permitted
by the memory limitations of the flash cluster nodes used for the computations.
The initial grid blocks thus have 48 cells in $x$ and $z$,  and 72 or  
96 in $y$. With 4 levels of refinement, i.e., 3 levels beyond the initial grid 
blocks, there are $2^3 \times 48 = 384$ cells initially in $x$ and $z$, yielding 
an initial cell size of $10^{15}$ cm $\approx 3 \times 10^{-4}$ pc in all three 
coordinate directions (the cells are cubical). With the maximum of 8 levels of 
refinement used, the minimum cell size drops by a factor of 16 to 
$\approx 6 \times 10^{13}$ cm $\approx 2 \times 10^{-5}$ pc.

\section{Results}

 Models A through I varied primarily in the assumed value of $\rho_{cr}$, in order
to assess the best choice for approximating the results obtained for a
full radiative transfer solution for a collapsing dense cloud core.
Vaytet et al. (2013) found that the evolution of the central temperature of a 
collapsing solar-mass cloud did not differ greatly between 
multigrid radiative transfer simulations with 20 different radiation frequency
bands and grey simulations with a single effective frequency band (cf. their
Figure 7), that is, at a given central density, the central temperatures
in the two simulations agreed to within about 10\%. They found that central
temperatures $T_c$ began to rise above the initial cloud temperature of 10 K for
central densities above $\rho_c \sim 10^{-13}$ g cm$^{-3}$, as expected 
(e.g., Boss 1980), with $T_c \approx 26$ K at $\rho_c = 10^{-12}$ g cm$^{-3}$.
In fact, the dependence of $T_c$ on $\rho_c$ in Figure 7 of Vaytet et al. (2013)
is almost exactly $T_c \propto \rho_c^{2/5}$, as expected for a polytropic gas 
composed primarily of molecular hydrogen with $\gamma = 7/5$.

 Model A was calculated with $\rho_{cr} = 10^{-14}$ g cm$^{-3}$, resulting in a 
rotationally flattened disk orbiting around a dense central protostar, but with a
central temperature $T_c \approx 300$ K when $\rho_c \sim 10^{-13}$ g cm$^{-3}$,
much too hot compared to the Vaytet et al. (2013) results. Similarly, models
B and C, both with $\rho_{cr} = 10^{-13}$ g cm$^{-3}$ but with different initial
rotation rates, produced disks with $T_c \approx 200$ K and $\approx 100$ K,
respectively, when $\rho_c \sim 10^{-13}$ g cm$^{-3}$, again too hot.
Models D and E, both with $\rho_{cr} = 3 \times 10^{-13}$ g cm$^{-3}$, 
were also too hot, with $T_c \approx 150$ K at $\rho_c \sim 10^{-13}$ g cm$^{-3}$.
On the other hand, models H and I, both with $\rho_{cr} = 10^{-12}$ g cm$^{-3}$, 
were much too cool, with $T_c \approx 10$ K when $\rho_c \sim 10^{-12}$ g cm$^{-3}$,
considerably lower than the Vaytet et al. (2013) value of $\approx$ 26 K at that
density. As a result, yet another two models were calculated, models F and G,
with $\rho_{cr} = 5 \times 10^{-13}$ g cm$^{-3}$, and these models finally
yielded a reasonable central temperature of $\approx$ 20 K when 
$\rho_c \sim 3 \times 10^{-13}$ g cm$^{-3}$. As a result, 
$\rho_{cr} = 5 \times 10^{-13}$ g cm$^{-3}$ was adopted as the best approximation
for reproducing the Vaytet et al. (2013) results.

 While these models were able to calibrate the choice of $\rho_{cr}$, it became
clear that a second set of models would need to be calculated with a longer box
in the $y$ direction. This is because the use of the new handling of the thermodynamics
in the optically thick central protostar meant that the calculations could be
carried forward farther in time than in the past, beyond when the collapsing
protostars and disks, accelerated by the shock to speeds of several km/sec, reached
the ends of the computational volume in models A through I. 
As a result, models J through O were calculated with the total computational
volume expanded by about a third along the direction of the shock wave propagation. 
In order to preserve the same grid resolution, this meant that the number of
computational blocks in the $\hat y$ direction had to be increased from 9 to 12
(Table 1). Each of these models required about two months of time to run while
using 88 cores on the DTM flash cluster, for a total of about 130K core-hours.

 Model L is indicative of the results for all six models with the extended 
computational grid. Figure 1 shows that the model L cloud has collapsed 
and formed a well-resolved central protostar orbited by a rotating, 
flattened protostellar disk, aligned perpendicular to the rotation axis 
($\hat y$) of the initial target cloud (see Boss \& Keiser 2015 for the 
results of models when the initial rotation axis of the cloud is perpendicular 
to the direction of motion of the shock front). 
The disk in Figure 1 is very similar in size to the corresponding 
disk seen in Figure 2 of Boss \& Keiser 2015. The disk mass is about 
0.3 $M_\odot$, while the central protostar has a mass of about 0.5 $M_\odot$,
implying that this is still a protostellar disk.
The maximum density in the protostar has risen to $\sim 10^{-11}$ g cm$^{-3}$,
well within the nonisothermal regime, and continues to rise slowly. Figure 2
shows the temperature distribution for model L at the same time as Figure 1.
The highest temperatures (up to 1000 K) occur in the shock front, while the
maximum temperature at the center of the protostar is $\sim$ 750 K. This
central temperature is considerably higher than that expected at this central
density, based on the spherically symmetric collapse models of Vaytet et al.
(2013), where the relevant central temperature was only about 56 K. Evidently
the FLASH 2.5 implementation of the
simple polytropic approximation fails to be valid at such high densities.
Even though the model could be computed farther in time than shown in 
Figures 1 and 2 by extending the computational grid even further in the 
$- \hat y$ direction, the limited validity of the polytropic approximation 
argues against such a continuation. Clearly then these models provide a
bounding model compared to our previous models, where the central protostellar
temperature remained fixed at 10 K as the nonisothermal regime was approached
and entered: these models show what happens when the protostar eventually 
becomes hotter than it should in a proper calculation with radiative transfer. 
In either case, the shock wave is able to trigger the formation and sustained
self-gravitational collapse of a central protostar, accompanied by the 
formation of a rotating protostellar disk. The models with higher initial
rotation rates (models K and M) formed disks roughly twice as wide as in model L, 
and the models with random noise in the initial density distribution (models
N and O) evolved very much like the corresponding models without any initial 
noise.

 The second requirement for the success of the supernova triggering and 
injection scenario is to examine the injection of material originally
residing in the shock front into the collapsing protostar and disk. 
Figure 3 shows that this requirement has been met as well in these
nonisothermal regime models. Figure 3 shows that the color field, which
is derived from the shock front, where it originally had unit space density, 
is distributed throughout the protostar, disk, and surrounding envelope. 
The color field space density is on the order of 0.05 units, 
meaning it has been diluted
by about a factor of 20 compared to its original space density in the 
shock front. Interestingly, the color field is significantly smaller inside 
the central protostar than in the surrounding envelope, implying that the
outer protoplanetary disk should be endowed with a higher space density
of SLRIs than the central protostar and the innermost disk, as suggested by 
early 2D models by Foster \& Boss (1996, 1997).

 Figures 4, 5, and 6 display the cloud density, temperature, and color 
fields for model L at the same time as in Figures 1, 2, and 3, plotted
now in the midplane of the edge-on disk evident in the first three figures.
The central protostar is seen clearly in Figure 4, along with a large scale,
$\sim$ 200 AU radius, swirling disk in the process of forming and
reaching quasi-equilibrium. Figure 5 shows that while the central
protostar has heated appreciably, the disk is largely isothermal at
10 K, as expected, as this disk is composed of gas well below the assumed
nonisothermal regime critical density $\rho_{cr} = 5 \times 10^{-13}$ 
g cm$^{-3}$ for model L. The contours of the hot shock front gas in
Figure 5 give an indication of the remnants of the Rayleigh-Taylor fingers
responsible for shock front injection. Figure 6 again shows that the
central protostar does not receive as large a dose of SLRIs as the forming
disk, or the envelope material that is on the periphery of the disk.
Figures 3 and 6 imply that SLRIs doses could vary by as much as 20\%
between the central and outer regions of this protostar and disk system.

 In order to test the sensitivity of the above results to the choice
of $\rho_{cr}$, we now consider model J, which is identical to model L,
except for having $\rho_{cr} = 10^{-12}$ g cm$^{-3}$ instead of
$\rho_{cr} = 5 \times 10^{-13}$ g cm$^{-3}$. Figures 7 and 8 depict
the color field distributions in the same two projections as for
model L in Figures 3 and 6, respectively. It can be seen by comparison
that the color field distributions are quite similar overall, differing
only in the fine detail. Evidently the precise choice of $\rho_{cr}$
has little effect on the outcome of the collapse, and in particular
on the degree and extent of the injection of shock front SLRIs into
the resulting protostar and disk system. 

 Figures 3 and 7 show that
the space density of the color field inside the collapsing protostar and
disk is $\approx 0.046$ (dimensionless units) for both models. 
Model 40-200-0.1-14 of Boss \& Keiser (2014) shows that for the same
model parameters, but without the nonisothermal heating, the color field in the
collapsing region (their Figure 6) is $\approx 0.025$. Given that the present
models have been advanced farther in time as a result of the nonisothermal
handling of the densest regions, to densities roughly ten times higher,
the new models show that the injection efficiency tends to increase with
time. As noted by Boss \& Keiser (2015), injection efficiencies also tend to
increase along with the amount of mass contained in the collapsing region.
This is to be expected by noting that the color field surrounding and
falling onto the disks seen in Figure 3 and 7 is largely more enriched
in SLRIs than the material already injected into the disk, implying
that the amount of injected SLRIs should continue to increase as the
collapse proceeds, at least to a certain extent. Defining the injection
efficiency factor $f_i$ as the fraction of the incident shock wave 
material that is injected into the collapsing cloud core, leads to an 
estimate of $f_i \approx 0.045$ for models L and J, somewhat higher
than the value of $f_i \approx 0.034$ for the comparable isothermal
model from Boss \& Keiser (2014). Given that the values of $f_i$
found by Boss \& Keiser (2014) led to dilution factors at the low
end of the range estimated by Takigawa et al. (2008) to be necessary
to explain the initial SLRI abundances when derived from a core collapse
SNe, the evidence for an increase in $f_i$ found in the present models,
as well as those of Boss \& Keiser (2015), lends further support in favor 
of the Type II supernova triggering and injection scenario for the solar
system.

\section{Dilution Factors}

 The geometric dilution factor $D_g$ may be defined as the fraction
of the total amount of mass in the SNR that is incident upon a target
cloud of radius $r$ lying at a distance $R$ from the stellar remnant and
is given by $D_g = r^2 / 4 R^2$. For our target cloud with $r = 0.053$ pc,
and assuming a nominal distance $R = 5$ pc (e.g., the radius of the 
Cygnus loop SNR, Blair et al. 1999), $D_g \approx 3 \times 10^{-5}$.
$D_g$ is the maximum fraction of the total amount of SN-ejected material 
that can be injected into the target cloud, assuming 100\% injection 
efficiency. For a 25 $M_\odot$ pre-supernova star and a neutron star
remnant, then, at most $\approx 7.2 \times 10^{-4} M_\odot$ of SN-derived
matter can be injected into the target cloud, based on geometric
dilution alone, assuming a nominal distance of 5 pc.

 Boss \& Keiser (2014) noted that geometric dilution is not the dilution
factor used their previous studies and in cosmochemical estimates of initial
SLRI abundances (e.g., Takigawa et al. 2008). This latter dilution factor $D$
is defined as the ratio of the amount of mass derived from the supernova
to the amount of mass derived from the target cloud. A SNR
with an initial speed of $\sim 4000$ km s$^{-1}$ must
be slowed down considerably by snowplowing the intervening interstellar
medium (ISM) in order to reach speeds of $\sim 40$ km s$^{-1}$ that 
do not result in cloud shredding (e.g., Boss et al. 2010). The factor
$\beta$ is defined as the ratio of shock front mass originating in the SN 
to the mass swept-up in the intervening ISM, leading to $\beta \approx 0.01$ 
for a SNR shock slowed down by a factor of $\sim$ 100 (e.g., Boss \& Keiser 2012).
For the models considered here, the amount of mass in the shock front that is
incident on the target cloud is 0.31 $M_\odot$, so that the amount
of shock front mass from the SN that is injected into the cloud 
is 0.31 $f_i \beta M_\odot$. For a solar-mass protostar, then,
$D \approx 0.31 f_i \beta$. Given that models L and J led to $f_i \approx 0.045$,
and taking $\beta \approx 0.01$, we get $D \approx 1.4 \times 10^{-4}$.
This dilution factor is on the lower end of cosmochemical estimates
of $D$ in the range of $1.3 \times 10^{-4}$ to $1.9 \times 10^{-3}$
for pre-supernova stars with masses of 25 $M_\odot$ and 20 $M_\odot$,
respectively (Takigawa et al. 2008).

 If $D \approx 1.4 \times 10^{-4}$, then for a solar-mass total system
mass, the total amount of SN material must be $\approx 1.4 \times 10^{-4}$
$M_\odot$, which is 5.1 times smaller than the maximum permitted by geometric
dilution of $\approx 7.2 \times 10^{-4} M_\odot$, for the parameters assumed
in these estimates. Given that $f_i \approx 0.045$, this implies that
geometric dilution would limit the amount of SN material to be at most
$\approx 3.2 \times 10^{-5} M_\odot$, i.e., about 4.4 times less
is required for $D \approx 1.4 \times 10^{-4}$. This implies that
in order for this $D$ value to agree with $D_g$, the target cloud
must have been closer than 5 pc away, e.g., at a distance of $\sim 2.4$ pc,
in order to increase $D_g$ by a factor of 4.4. A consistent result
for the injection efficiency based on estimates of both $D$ and $D_g$,
at least for a 25 $M_\odot$ pre-supernova star, might then require that the 
target cloud lie about 2.4 pc from the pre-supernova star, or perhaps that
the pre-supernova star was more massive than 25 $M_\odot$.

 The W44 SNR is observed to be compressing an adjacent molecular cloud
at a phase when the W44 SNR has a radius of about 11 pc (Reach et al. 2005),
considerably more distant than the 2.4 pc suggested above. The Cas A SNR 
has a radius of only about 1.5 pc (e.g., Krause et al. 2008), but is not 
close to a GMC region. A consistent result may require a close
association between a target molecular cloud core and a massive,
pre-supernova star. Oullette et al. (2010), e.g., proposed 
having a SN inject SLRIs into an already formed protoplanetary disk 
lying only 0.3 pc away. Such a close association is supported by
observations of the Eagle Nebula (M16), where dense cloud cores are
being irradiated by a cluster of O stars (some as massive as 80 $M_\odot$)
only about 1 pc away.

 Yet another plausible possible solution to the consistency problem
involves the fact that SN ejecta are quite clumpy, with large variations
in the abundances of SLRIs such as $^{44}$Ti (factors of 4 or more)
having been observed in the Cas A SNR (Grefenstette et al. 2014), 
implying that core collapse SN explosions can be highly asymmetric.
Hence discrepancies on the order of factors of 4 or more in SLRI 
abundances could be simply explained as a result of the presolar cloud
having been triggered into collapse by a particularly SLRI-rich, dense 
portion of a SNR. Clearly there are several plausible means for 
reconciling differing estimates of $D_g$ and $D$.

\section{Discussion}

 Sahijpal \& Goswami (1998) proposed that the absence of evidence for live 
$^{26}$Al in the rare FUN (Fractionation Unknown Nuclear) CAIs was a 
result of the formation of the FUN inclusions prior to normal CAIs from 
collapsing gas and dust interior to the outer collapsing regions containing 
the freshly injected SN SLRIs, as had been suggested by the early shock 
triggered collapse models of Foster \& Boss (1996, 1997) and as can be
seen to a lesser extent in Figures 3 and 7 for the present models. 
MacPherson \& Boss (2011) suggested instead that the FUN inclusions
may have originated in a SLRI-poor young stellar object (YSO), ejected
by the YSO's bipolar outflow, and transported to a presolar cloud core
in the same star-forming region.
Kuffmeier et al. (2016) have argued against both of these suggestions,
in the latter case on the basis that the inclusions would have to come
from more than $\sim$ 0.25 pc away, which they judged to be unrealistic.
With regard to the former suggestion, Kuffmeier et al. (2016)
included as a second step zoom-in calculations
of the GMC regions where eleven low mass stars formed, represented
in the GMC code as sink particles, with a focus on the SLRI levels
in the immediate vicinity (50 AU or 1000 AU) of the sink particles. The SLRI
ratios were uniformly homogeneous at varied levels, implying there
should be no variation in the SLRI ratios during the accretion process.
However, Kuffmeier et al. (2016) did find SLRI ratio variations during
the late phases, $\sim$ 0.1 Myr after sink particle formation. For one
star, the $^{26}$Al/$^{27}$Al ratio of accreted material increased
by a factor of 2 after 0.16 Myr, but this heterogeneity was attributed 
to the fact that the sink particle had already accreted most of the
gas in its vicinity, allowing more distant gas to be accreted. 
In fact, because shock-triggered star formation does not occur in
the Kuffmeier et al. (2016) models, there is little mixing between 
the hot, low density GMC regions with fresh SLRIs, and the cool, high density
gas in pre-stellar cores. The models presented here, on the other hand,
are specifically intended to explore RT injection of fresh SLRIs carried
by SN shocks into pre-stellar cores, a process not studied by the
Kuffmeier et al. (2016) simulation. Kuffmeier et al. (2017) presented
further zoom-in analysis of the Vasileiadis et al. (2013) and Kuffmeier 
et al. (2016) GMC simulation, showing evidence for strongly
heterogeneous accretion of gas and dust in space and time, though
they did not address the question of the possible implications of
this heterogeneity for the FUN inclusions.

 Larsen et al. (2016) presented analyses of achondritic
meteorites, and a few chondrites, which they interpreted as evidence
of initial spatial heterogeneity of the $^{26}$Al/$^{27}$Al ratio.
Larsen et al. (2016) proposed that $^{26}$Al was fractionated by thermal
processing of the carrier dust grains into gas and dust, which
thereafter separated and retained different $^{26}$Al/$^{27}$Al ratios, 
resulting in a solar nebula with a heterogeneous $^{26}$Al distribution, with
initial $^{26}$Al/$^{27}$Al ratios varying from $\sim 1 - 2 \times 10^{-5}$
in the inner nebula to $\ge 2.7 \times 10^{-5}$ in the outer nebula, 
compared to an initial canonical ratio of $5.25 \times 10^{-5}$ in CAIs. This
interpretation relies on the mechanism suggested by Trinquier et al. (2009)
and Schiller et al. (2015), who claimed that newly-synthesized SLRIs
would be carried by grains that are more thermally labile than the
older grains carrying more ancient nucleosynthetic products, resulting
in the nebular gas being enriched in $^{26}$Al compared to the 
dust. Kuffmeier et al. (2016) appealed to this same explanation to account
for the lack of evidence for live $^{26}$Al in FUN inclusions, and extended
this argument to the case of $^{60}$Fe, which they found to be highly
overabundant in their sink particles compared to the solar system's
initial inventory. They argued that the carrier phase of the $^{60}$Fe
must be even more volatile than that for the $^{26}$Al, so that both
SLRIs could be selectively removed from the dust grains that would go
on to form the CAIs. However, given that nearly all SNR dust grains are 
expected to be sub-micron in size (Bocchio et al. 2016), even if some of 
these grains were thermally evaporated in the infalling presolar cloud, 
or in the solar nebula, and others were not, to first order these small 
grains will be transported and mixed along with the gas 
(e.g., Bate \& Loren-Aguilar 2017),  
eliminating the chance of any large-scale isotopic fractionation that
might survive in the solar nebula. Solid particles need to grow to roughly
cm-size before appreciable decoupling of their orbital motions from that
of the disk gas occurs (e.g., Boss 2015).

\section{Conclusions}

 These 3D triggering and injection models have investigated the effects 
of the loss of molecular line cooling once the target cloud becomes 
optically thick at densities above $\sim 10^{-13}$ g cm$^{-3}$, when the 
collapsing regions begin to heat above 10 K, but continue their collapse 
toward the formation of the first protostellar core at a central density of 
$\sim 10^{-10}$ g cm$^{-3}$ (e.g., Boss \& Yorke 1995). The models show
that the SLRI injection efficiencies ($f_i$) tend to increase as the amount
of collapsing matter increases, and as the central density approaches
that of the formation of the first protostellar core. These new values of
$f_i \approx 0.045$, combined with previously estimated dilution factors for 
core collapse SNRs (Boss \& Keiser 2014), produce an improved agreement
with the predicted dilution factors needed to explain the solar system’s 
SLRIs (Takigawa et al. 2008) in the context of the supernova triggering 
and injection scenario. Ideally, these models will be able to tie a SNR 
producing similar injection efficiencies to a specific core-collapse SNe 
nucleosynthesis model, rather than having to invoke a generic GMC
scenario (e.g., Kuffmeier et al. 2016) involving a mixture of SNe
contributions and massive star winds (e.g., Young 2014, 2016), 
over a time period of several Myr, to explain the origin of the 
solar system SLRIs.

 The FLASH 4.3 code allows the use of sink particles 
to represent the central protostars, sidestepping the issue of resolving
the thermal evolution of the growing protostar beyond the first core.
Given the evident need to further explore the implications for the triggering
and injection scenario, a set of FLASH 4.3 calculations with sink particles 
is presently underway on the flash cluster, and the results will be 
presented in a future paper.

\acknowledgments

 Sandra A. Keiser was to be the co-author on this paper, but she passed
away unexpectedly before this work was finished. I dedicate this paper to 
Sandy, in honor of her decades of superb computational support for the DTM 
astronomy group. I thank Michael Acierno for his help with the DTM flash 
cluster, where the calculations were performed, and the referee for a 
number of constructive comments. The software used in this 
work was in large part developed by the DOE-supported ASC/Alliances Center 
for Astrophysical Thermonuclear Flashes at the University of Chicago.

\clearpage
\begin{deluxetable}{lccccccc}
\tablecaption{Initial conditions for the models: initial
target cloud rotation rates ($\Omega_i$, in rad s$^{-1}$), 
critical density for polytropic pressure law ($\rho_{cr}$, in g cm$^{-3}$),
number of initial grid blocks in $\hat y$ direction ($N_y$),
minimum $y$ grid value ($y_{min}$, in cm), initial number of levels
of refinement ($N_i$), final number of levels of refinement ($N_f$),
and whether the initial ambient density
distributions were uniform or had random noise. \label{tbl-1}}
\tablewidth{0pt}
\tablehead{\colhead{Model} 
& \colhead{$\Omega_i$} 
& \colhead{$\rho_{cr}$} 
& \colhead{$N_y$}
& \colhead{$y_{min}$} 
& \colhead{$N_i$}
& \colhead{$N_f$}
& density }
\startdata

A & $10^{-14}$ & $ 10^{-14}$         &   9  & 0.0                   & 4 & 6 & uniform \\

B & $10^{-14}$ & $ 10^{-13}$         &   9  & 0.0                   & 4 & 6 & uniform \\

C & $10^{-13}$ & $ 10^{-13}$         &   9  & 0.0                   & 4 & 5 & uniform \\

D & $10^{-14}$ & $3 \times 10^{-13}$ &   9  & 0.0                   & 4 & 5 & uniform \\

E & $10^{-13}$ & $3 \times 10^{-13}$ &   9  & 0.0                   & 4 & 5 & uniform \\

F & $10^{-14}$ & $5 \times 10^{-13}$ &   9  & 0.0                   & 4 & 8 & uniform \\

G & $10^{-13}$ & $5 \times 10^{-13}$ &   9  & 0.0                   & 4 & 7 & uniform \\

H & $10^{-14}$ & $ 10^{-12}$         &   9  & 0.0                   & 4 & 5 & uniform \\

I & $10^{-13}$ & $ 10^{-12}$         &   9  & 0.0                   & 4 & 5 & uniform \\

J & $10^{-14}$ & $ 10^{-12}$         &  12  & $-2.0 \times 10^{17}$ & 4 & 8 & uniform \\

K & $10^{-13}$ & $ 10^{-12}$         &  12  & $-2.0 \times 10^{17}$ & 4 & 7 & uniform \\

L & $10^{-14}$ & $5 \times 10^{-13}$ &  12  & $-2.0 \times 10^{17}$ & 4 & 8 & uniform \\

M & $10^{-13}$ & $5 \times 10^{-13}$ &  12  & $-2.0 \times 10^{17}$ & 4 & 6 & uniform \\

N & $10^{-14}$ & $5 \times 10^{-13}$ &  12  & $-2.0 \times 10^{17}$ & 3 & 7 & noise   \\
 
O & $10^{-13}$ & $5 \times 10^{-13}$ &  12  & $-2.0 \times 10^{17}$ & 3 & 8 & noise   \\

\enddata
\end{deluxetable}

\begin{figure}
\vspace{-1.0in}
\includegraphics[scale=.80,angle=90]{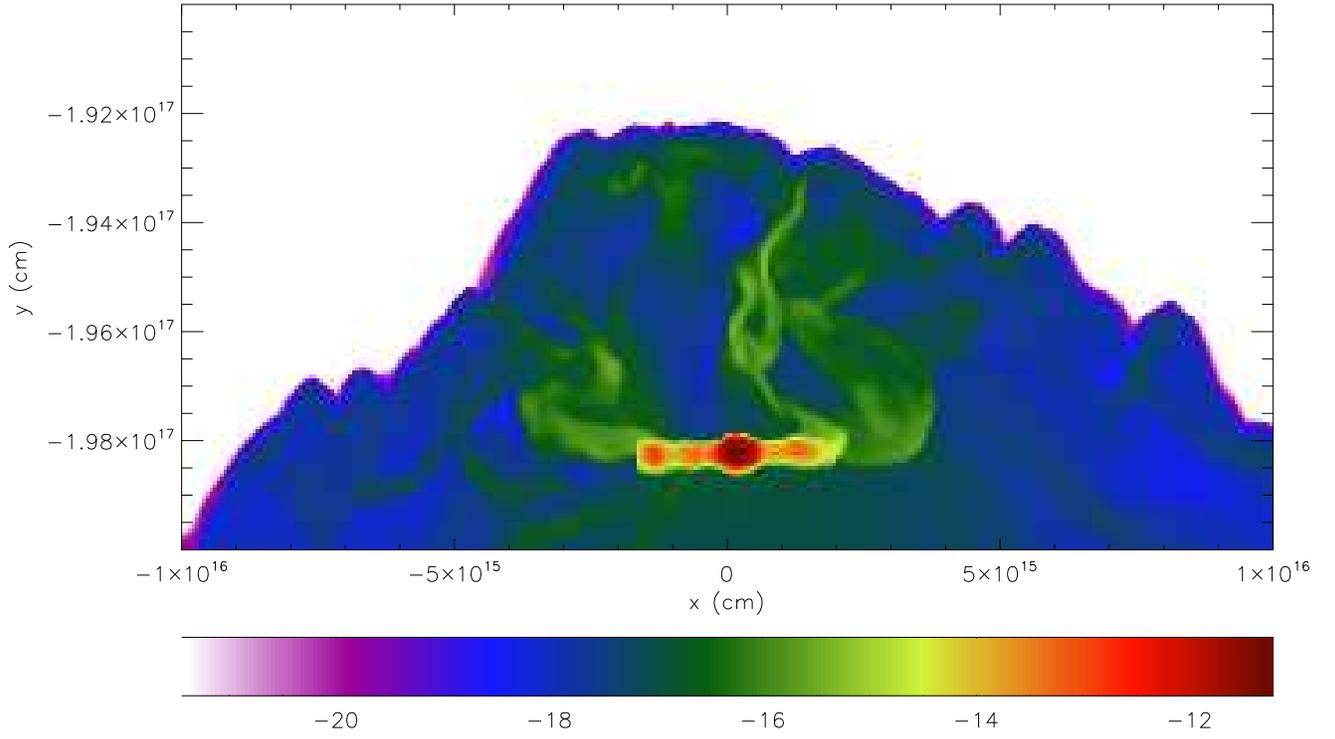}
\vspace{-0.5in}
\caption{Model L log density cross-section ($z$ = 0) at 0.0869 Myr.}
\end{figure}
\clearpage

\begin{figure}
\vspace{-1.0in}
\includegraphics[scale=.80,angle=90]{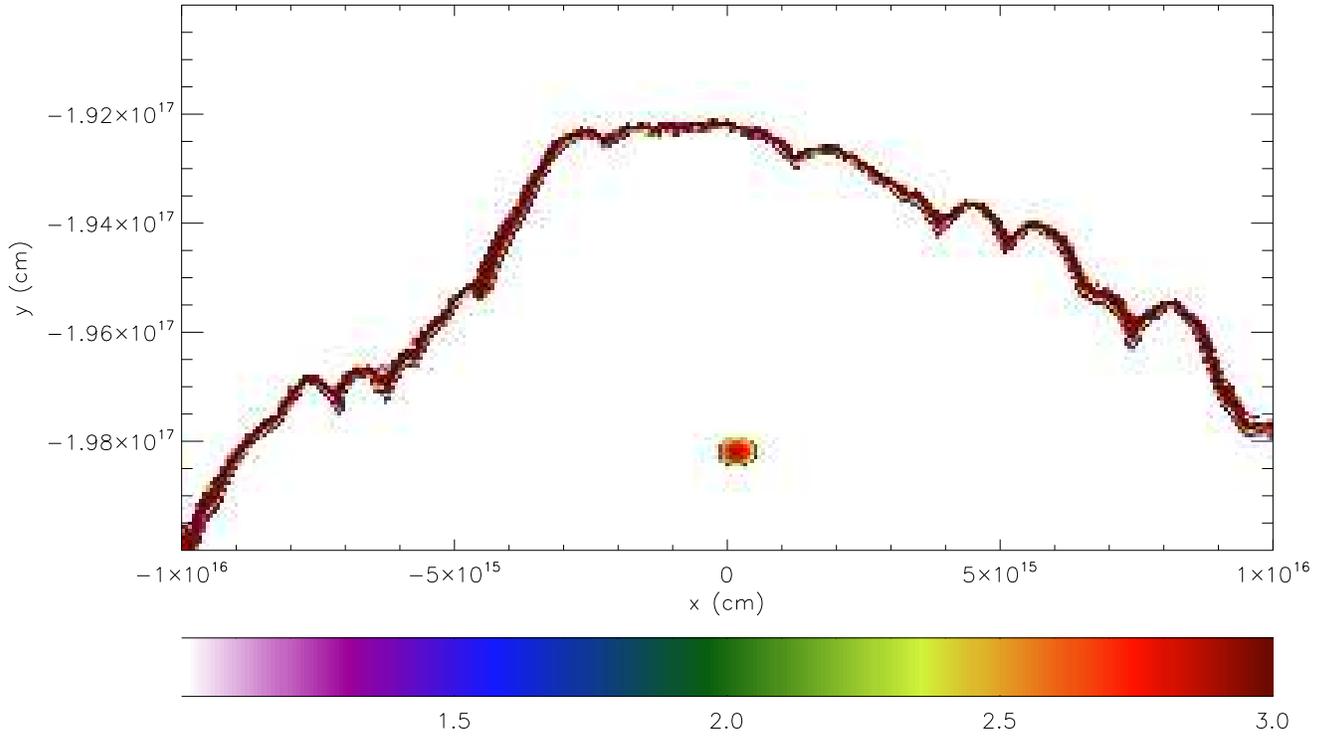}
\vspace{-0.5in}
\caption{Model L log temperature cross-section ($z$ = 0) at 0.0869 Myr.
The corrugated structure is a reflection of the Rayleigh-Taylor fingers
responsible for shock wave matter injection.}
\end{figure}
\clearpage

\begin{figure}
\vspace{-1.0in}
\includegraphics[scale=.80,angle=90]{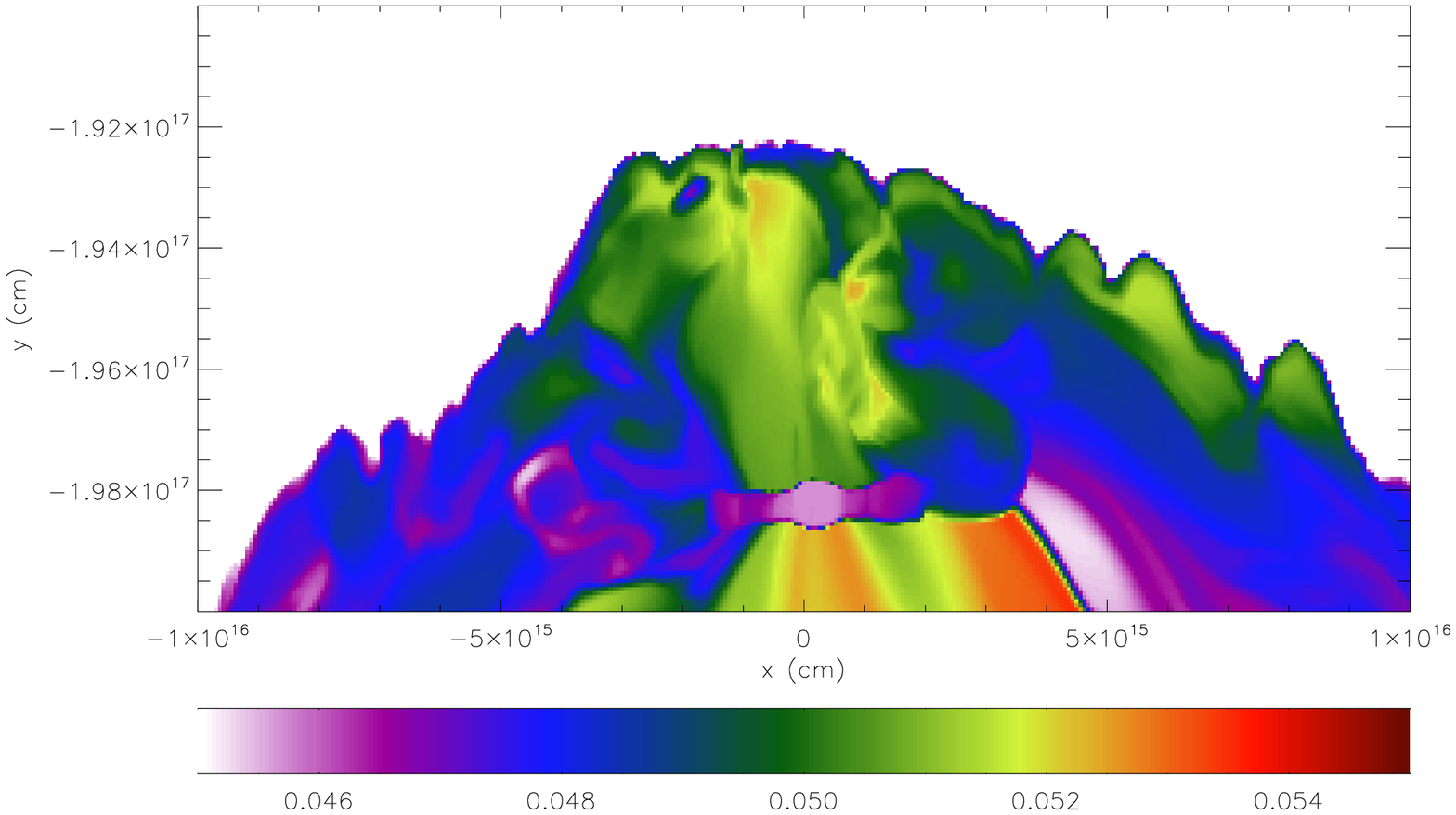}
\vspace{-0.5in}
\caption{Model L color field (shock front matter) cross-section ($z$ = 0) at 0.0869 Myr.}
\end{figure}
\clearpage

\begin{figure}
\vspace{-1.0in}
\includegraphics[scale=.60,angle=90]{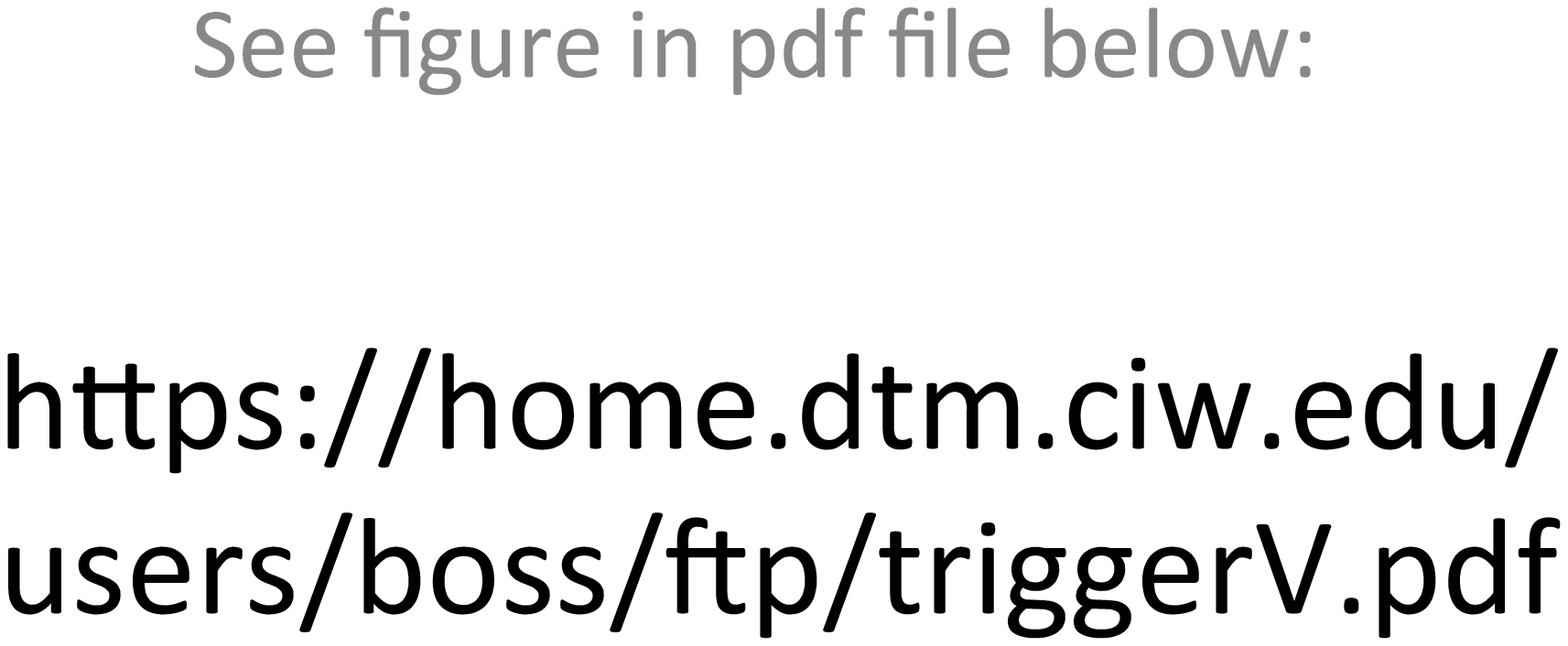}
\vspace{-0.5in}
\caption{Model L log density cross-section in midplane ($y = - 1.98 \times 10^{17}$ cm) 
of disk at 0.0869 Myr.}
\end{figure}
\clearpage

\begin{figure}
\vspace{-1.0in}
\includegraphics[scale=.60,angle=90]{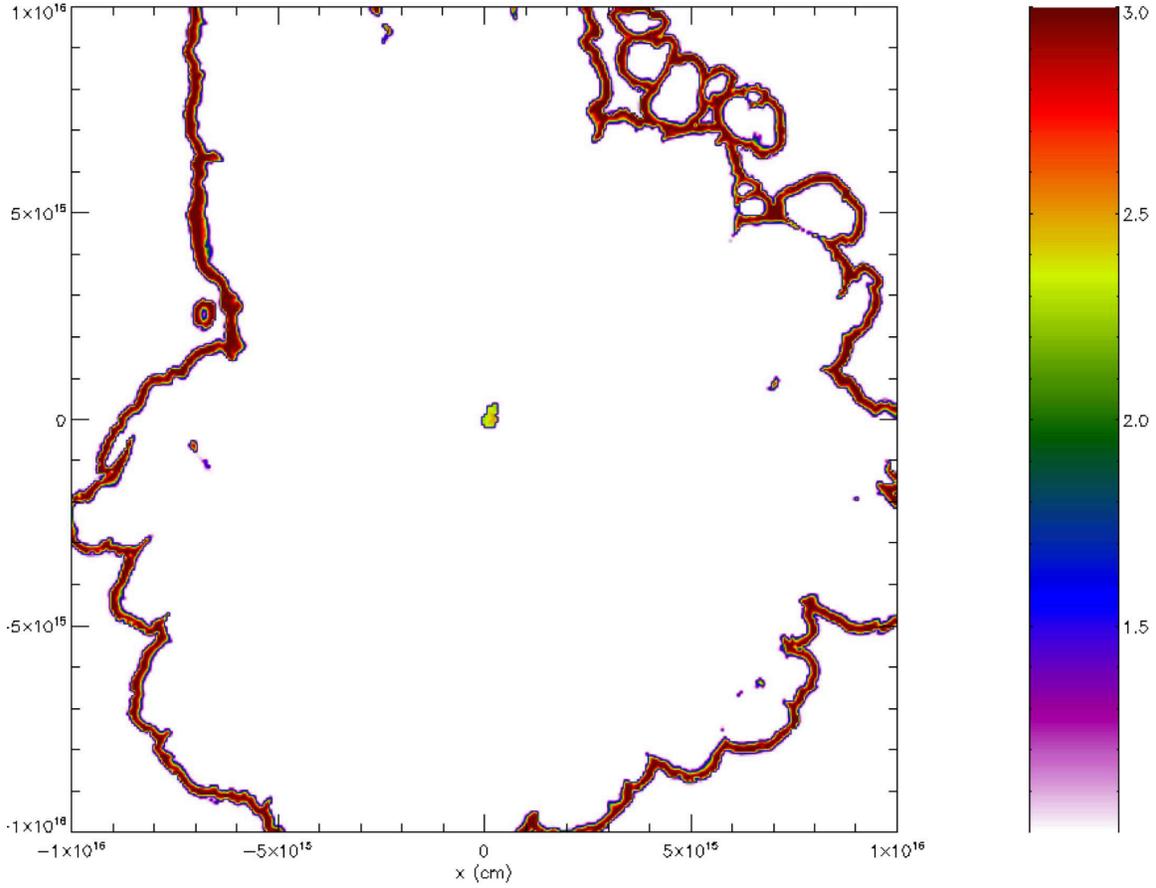}
\vspace{-0.5in}
\caption{Model L log temperature cross-section in midplane ($y = - 1.98 \times 
10^{17}$ cm) of disk at 0.0869 Myr.
The corrugated structure is a reflection of the Rayleigh-Taylor fingers
responsible for shock wave matter injection.}
\end{figure}
\clearpage

\begin{figure}
\vspace{-1.0in}
\includegraphics[scale=.60,angle=90]{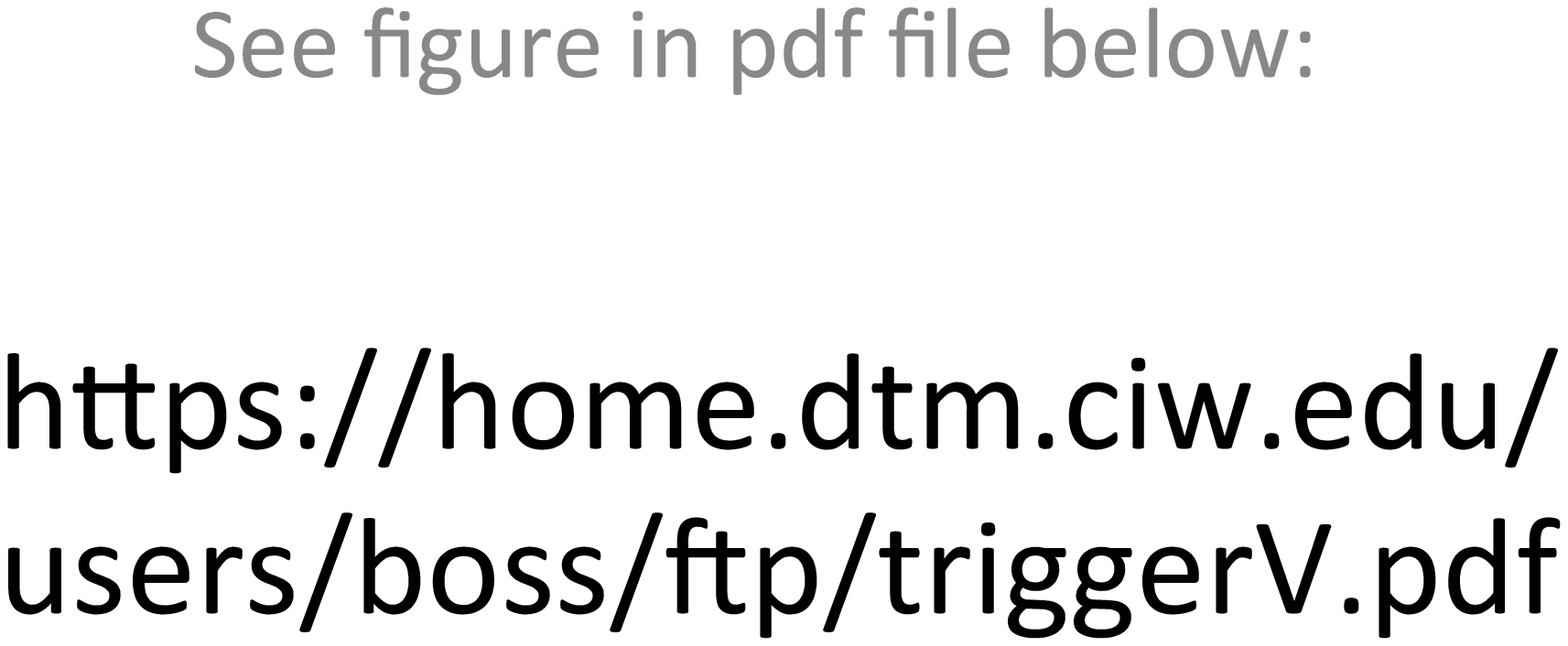}
\vspace{-0.5in}
\caption{Model L color field (shock front matter) cross-section in midplane 
($y = - 1.98 \times 10^{17}$ cm) of disk at 0.0869 Myr.}
\end{figure}
\clearpage

\begin{figure}
\vspace{-1.0in}
\includegraphics[scale=.80,angle=90]{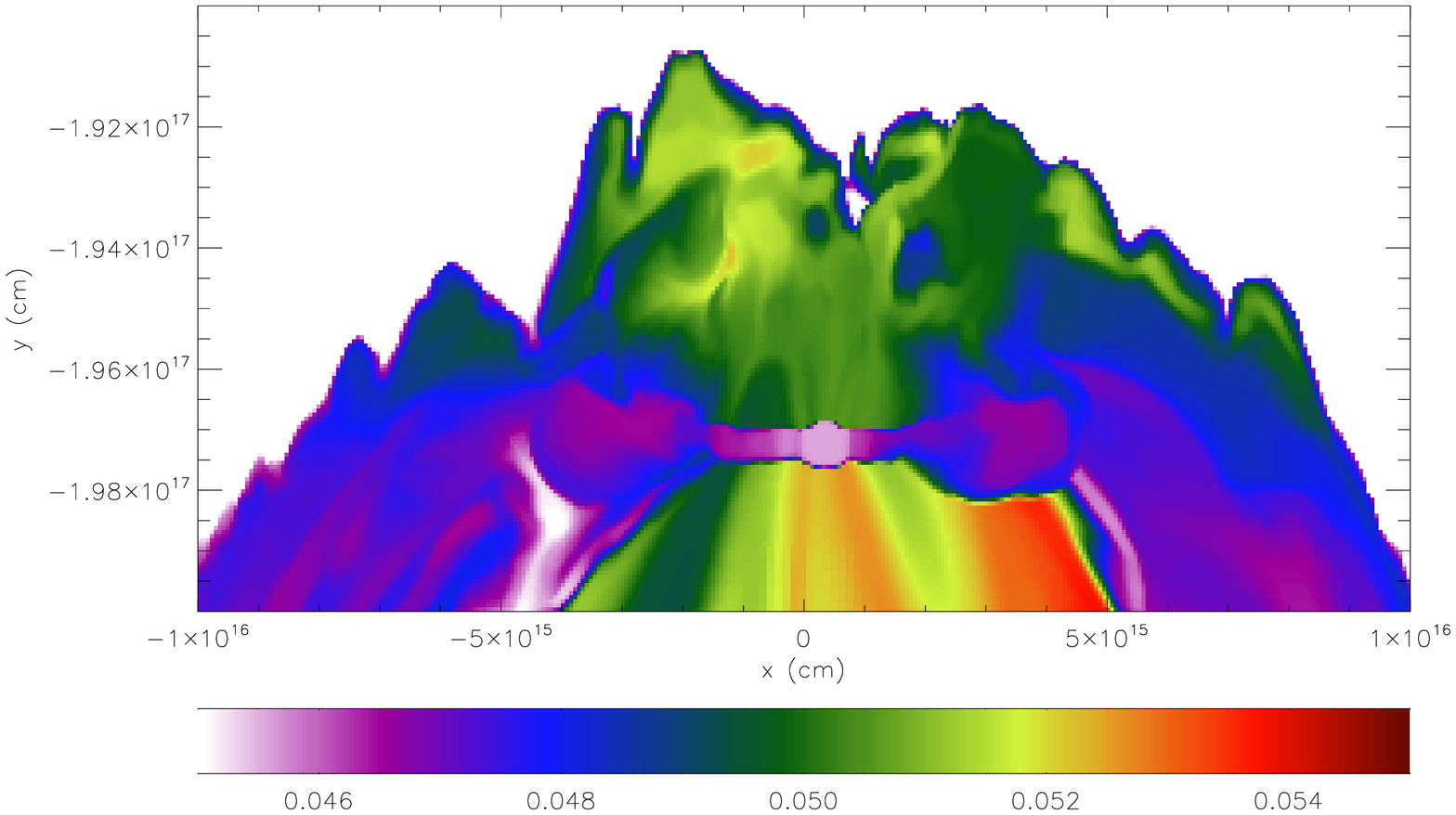}
\vspace{-0.5in}
\caption{Model J color field (shock front matter) cross-section ($z$ = 0) at 0.0868 Myr.}
\end{figure}
\clearpage

\begin{figure}
\vspace{-1.0in}
\includegraphics[scale=.60,angle=-90]{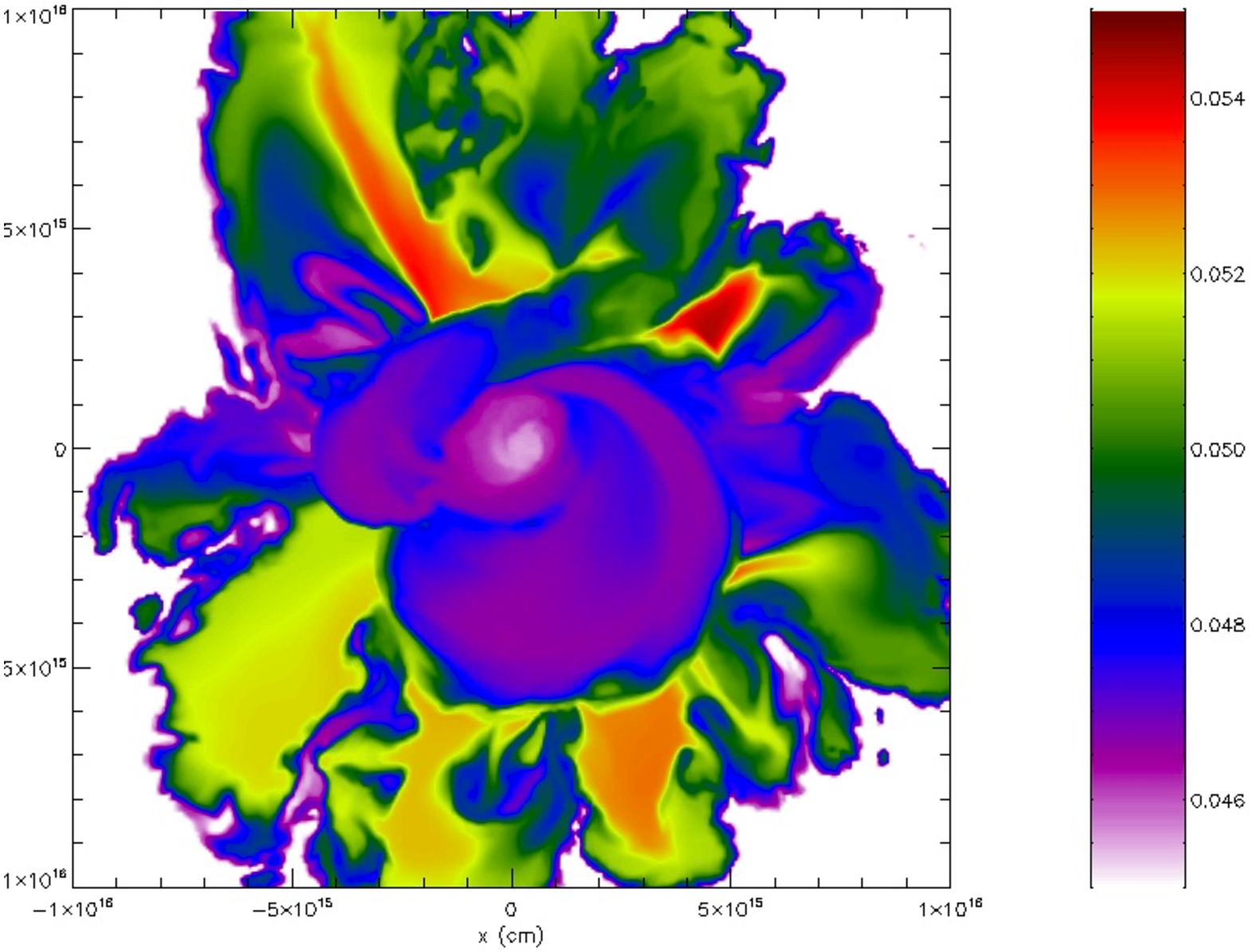}
\vspace{-0.5in}
\caption{Model J color field (shock front matter) cross-section in midplane 
($y = - 1.973 \times 10^{17}$ cm) of disk at 0.0868 Myr.}
\end{figure}
\clearpage

\clearpage


\begin{references}

\reference{r}
Balazs, L. G., Abraham, P., Kun, M., Kelemen, J., \& Toth, L. V.
2004, A\&A, 425, 133

\reference{r}
Bate, M. R., \& Loren-Aguilar, P. 2017, MNRAS, 465, 1089

\reference{r}
Banerjee, P., Qian, Y.-Z., Heger, A., \& Haxton, W. C. 2016, Nature
Communications, 7:13639

\reference{r}
Blair, W. P., Sankrit, R., Raymond, J. C., \& Long, K. S. 1999, 
AJ, 118, 942

\reference{r}
Bocchio, M., Marassi, S., Schneider, R., Bianchi, S., Limongi, M., \&
Chieffi, A., 2016, A\&A, 587, A157

\reference{r}
Boss, A. P. 1980, ApJ, 242, 699 

\reference{r}
Boss, A. P. 1995, ApJ, 439, 224

\reference{r}
Boss, A. P. 2015, ApJ, 807, 10

\reference{r}
Boss, A. P. 2016, in Handbook of Supernovae, eds. A. W. Alsabti, P. Murdin
(Switzerland: Springer)

\reference{r}
Boss, A. P., \& Keiser, S. A. 2010, ApJL, 717, L1

\reference{r}
Boss, A. P., \& Keiser, S. A. 2013, ApJ, 770, 51 

\reference{r}
Boss, A. P., \& Keiser, S. A. 2014, ApJ, 788, 20

\reference{r}
Boss, A. P., \& Keiser, S. A. 2015, ApJ, 809, 103

\reference{r}
Boss, A. P., \& Yorke, H. A. 1995, ApJ,  439, L55

\reference{r}
Boss, A. P., Ipatov, S. I., Keiser, S. A., Myhill, E. A., \& Vanhala, 
H. A. T. 2008, ApJ, 686, L119

\reference{r}
Boss, A. P., Keiser, S. A., Ipatov, S. I., Myhill, E. A., \& Vanhala, 
H. A. T. 2010, ApJ, 708, 1268

\reference{r}
Cameron, A. G. W., \& Truran, J. W. 1977, Icarus, 30, 447

\reference{r}
Chevalier, R. 1974, ApJ, 188, 501

\reference{r}
Falle, S. A. E. G., Vaidya, B., \& Hartquist, T. W. 2017, MNRAS, 465, 260

\reference{r}
Foster, P. N., \& Boss, A. P. 1996, ApJ, 468, 784 

\reference{r}
Foster, P. N., \& Boss, A. P. 1997, ApJ, 489, 346

\reference{r}
Fryxell, B., Olson, K., Ricker, P., et al. 2000, ApJS, 131, 273

\reference{r}
Goodson, M. D., Luebbers, I., Heitsch, F., \& Frazer, C. C. 2016, MNRAS, 462, 2777

\reference{r}
Grefenstette, B. W., et al. 2014, Nature, 506, 339

\reference{r}
Jura, M., Xu, S., \& Young, E. D. 2013, ApJL, 775, L41

\reference{r}
Krause, O., et al. 2008, Science, 320, 1195

\reference{r}
Kuffmeier, M., Mogensen, T. F., Haugbolle, T., Bizzarro, M., \& Nordlund, A. 2016,
ApJ, 826, 22

\reference{r}
Kuffmeier, M., Haugbolle, T., \& Nordlund, A. 2017, ApJ, in revision

\reference{r}
Larsen, K. K., Schiller, M., \& Bizzarro, M. 2016, Geochim. 
Cosmochim. Acta, 176, 295

\reference{r}
Lee T., Papanastassiou, D. A., \& Wasserburg, G. J. 1976, Geophys. Res. Lett., 3, 109

\reference{r}
Li, S., Frank, A., \& Blackman, E. G. 2014, MNRAS, 444, 2884

\reference{r}
Lichtenberg, T., Parker, R. J., \& Meyer, M. R. 2016, MNRAS, 462, 3979

\reference{r}
MacPherson, G. J., \& Boss, A. P. 2011, PNAS,108, 19152

\reference{r}
Mishra, R. K., \& Chaussidon, M. 2014, EPSL, 398, 90

\reference{r}
Mishra, R. K., \& Goswami, J. N. 2014, Geochim. Cosmochim. Acta, 132, 440

\reference{r}
Mishra, R. K., Marhas, K. K., \& Sameer 2016, EPSL, 436, 71

\reference{r}
Nicholson, R. B., \& Parker, R. J. 2017, MNRAS, 464, 4318

\reference{r}
Ouellette, N., Desch, S. J., \& Hester, J. J. 2007, ApJ, 662, 1268

\reference{r}
Ouellette, N., Desch, S. J., \& Hester, J. J. 2010, ApJ, 711, 597

\reference{r}
Parker, R. J., \& Dale, J. E. 2016, MNRAS, 456, 1066

\reference{r}
Reach, W. T., Rho, J., \& Jarrett, T. H. 2005, ApJ, 618, 297

\reference{r}
Sahijpal, S., \& Goswami, J. N. 1998, ApJL, 509, L137

\reference{r}
Schiller, M., Paton, C., \& Bizzarro, M. 2015, Geochim. Cosmochim. Acta,
149, 88

\reference{r}
Tachibana, S., Huss, G. R., Kita, N. T., Shimoda, G., \&
Morishita, Y. 2006, ApJ, 639, L87

\reference{r}
Takigawa, A., Miki, J., Tachibana, S., et al. 2008, ApJ, 688, 1382

\reference{r}
Tang, H., \& Dauphas, N. 2012, EPSL, 59, 248

\reference{r}
Telus, M., et al. 2016, Geochim. Cosmochim. Acta, 178, 87

\reference{r}
Telus, M., Huss, G. R., Nagashima, K., Ogliore, R. C., \& Tachibana, S.
2017, Geochim. Cosmochim. Acta, in revision

\reference{r}
Trinquier, A., et al. 2009, Science, 324, 374

\reference{r}
Vanhala, H. A. T., \& Boss, A. P. 2000, ApJ, 538, 911

\reference{r}
Vasileiadis, A., Nordlund, A., \& Bizzarro, M. 2013, ApJ, 769, L8

\reference{r}
Vaytet, N., et al. 2013, A\&A, 557, A90

\reference{r}


\end{references}
\end{document}